\documentclass[aps,pra,twocolumn,showpacs,superscriptaddress]{revtex4}

\usepackage{amsmath}
\usepackage{epsfig}
\usepackage{bm}
\usepackage{longtable}
\usepackage{graphicx}
\usepackage{graphics}

\begin{document}

\title{Coherent transport in homojunction between 
excitonic insulator and semimetal
}

\author{Massimo Rontani}
\affiliation{INFM National Research Center on nanoStructures
and bioSystems at Surfaces (S3), 
Via Campi 213/A, 41100 Modena, Italy}

\author{L. J. Sham}

\affiliation{Department of Physics, University of California San Diego,
La Jolla, California 92093-0319}

\date{\today}

\begin{abstract}
From the solution of a two-band model, we predict that the thermal and 
electrical transport across the junction of a semimetal and an 
excitonic insulator will exhibit high resistance behavior and
low entropy production at low temperatures, distinct from a junction 
of a semimetal and a normal semiconductor. This phenomenon, ascribed 
to the dissipationless exciton flow which dominates over the charge transport, 
is based on the much longer length scale of the 
change of the
effective interface potential for electron scattering due to the coherence 
of the condensate than in the normal state.
\end{abstract}

\pacs{71.35.Lk, 72.10.Fk, 73.40.Ns, 73.50.Lw}

\maketitle

Proposals that exciton condensation transforms a semimetal
(SM) into an {\em excitonic insulator} (EI) date back more than four
decades \cite{keldysh,review,Snoke}. Experiments on indirect gap
semiconductors \cite{exp} and on coupled quantum wells \cite{butov}
have shown unusual properties which were inferred as indications of
the exciton condensation. Earlier, the difficulties in distinguishing
between the EI and the ordinary dielectric were pointed 
out \cite{Keldyshmaligno}. Here we show that, {\em if} an EI exists, 
the larger coherence length scale of the exciton condensation allows 
the existence of a
homojunction between an SM state and an EI state whereas the smaller
length scale of the structural interface between a semimetal and a
normal semiconductor gives it the nature of a heterojunction. 
Across the homogeneous SM/EI junction, the current is composed of
two competing terms associated, respectively, with 
neutral excitons and with charge carriers. 
At small electrical bias and low
temperature, exciton flow dominates over
the free charges, increasing substantially the electrical
and thermal interface resistance. The rate of entropy production is
low due to the coherent and dissipationless character of the exciton 
flow, analogous to the case of a clean metal/superconductor (NS) 
junction.

Consider a model of two overlapping bands 
which form a semimetal as in the left panel of Fig.~1(a). 
A semiconductor (SC) may be formed either by a one-electron 
hybridization potential to a normal insulating state (NSC) or by a 
strong electron interaction between the two bands into an 
excitonic insulator. A normal semiconductor may, for example, be formed by 
changing an element in a compound for the semimetal and an 
excitonic insulator may be formed from the semimetal by strain or suitable 
alloying of the SM compound with a third element.
Since the hybridization potential and the exciton 
condensation order parameter, $\Delta$, 
contribute additively to the small band 
gaps formed between the two overlapping bands as in the right panel 
of Fig.~1(a), the two resulting states cannot be 
distinguished by spectroscopy. We propose as an experimental means 
of identifying the EI state the transport properties across a 
junction with a sharp interface made of a semimetal and a small gap 
semiconductor, which is either normal or an EI. The key difference is 
in the length scale of the variation of 
the effective interface potential which 
reflects or transmits the electron. In the SM/EI junction, a 
component of the effective potential is the position dependent 
electron-hole paring potential,
$\Delta(\mathbf{r})$, which decreases from the 
bulk value in the EI region to zero in the SM region [see 
Fig.~2(a)]. The length scale of the change is the coherence 
length in EI, much longer than the lattice constant. The finite order 
in the SM region is due to the proximity effect of the exciton 
condensate in EI. Since this system works best if the lattices of the 
two components are as similar as possible, we classify the junction 
as \textit{homogeneous}  (termed homojunction by convention). On the 
other hand, in the SM/NSC junction, the one-electron interface 
potential due to the change in hybridization has 
an abrupt variation on a length scale of the order of a few atomic layers,
thus considered in our context as a \textit{heterogeneous} 
junction \cite{Schottky}.
\begin{figure}
\centerline{\epsfig{file=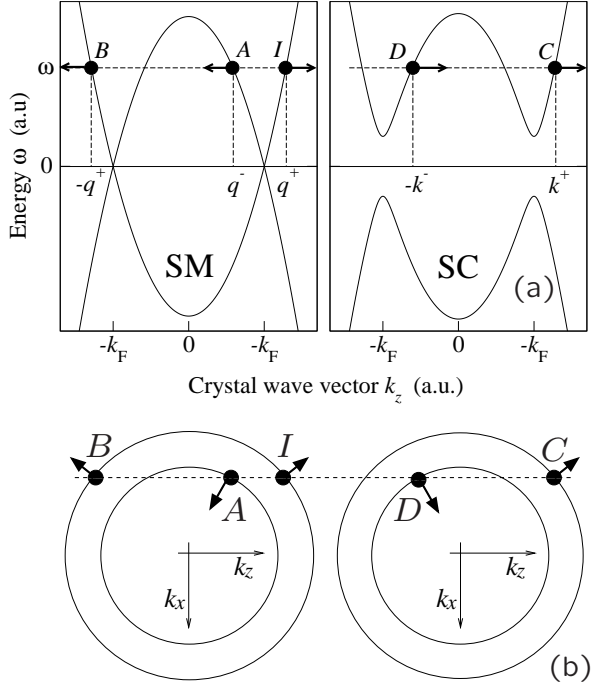,width=3.1in,,angle=0}}
\caption{Elastic scattering across the boundary between a semimetal
(SM, left) and a semiconductor (SC, right). (a) Dispersion of
the quasi-particle excitation energy $\omega$ vs.~wave vector.
The energy is referenced from the chemical potential ($\omega=0$).
The scattering channels of
an incident electron at point $I$ are indicated.
(b) Corresponding points on the electron and hole isoenergetic
surfaces together with the group velocity vectors.
\label{fig3}}
\end{figure}

An important consequence of the interface potential is that carriers 
coming from the bulk SM with energies slightly outside the SC gap 
have, say for the incident electron at $I$, two reflection channels, 
$A$ and $B$, and two transmission channels at $C$ and $D$ (see 
Fig.~1). If the energy lies within the gap, only the two 
reflection channels are possible. While the interface by breaking the 
lattice translational symmetry can in principle connect different 
parts of the Brillouin zone \cite{SN}, to express the results of 
our detailed model study of the interface scattering it is important 
to consider the regions of the wave vector space near the gaps
in the bulk SC as two valleys 
for the states with the same component of the wave vector parallel 
to the interface. In the SM/EI homojunction, the 
slowly varying electron-hole pairing potential causes the dominant 
scattering to be intravalley, i.e., from $I$ to $A$ and $C$. In the 
SM/NSC heterojunction, the rapid spatial variation of the interface 
potential causes strong intervalley reflection, i.e., from $I$ to 
$B$, and strong intravalley transmission, i.e., from $I$ to $C$. 
We have modeled the common physical features of the heterojunction, 
including the abrupt band edge discontinuity, the short-ranged 
interface potential, and the impurities at the interface (see Appendix). 
We shall now focus on the intravalley reflection arising out of 
the band mixing from the EI. 

For simplicity of exposition, we take the effective masses of the
two bands to be isotropic and equal to $m$ (Fig.~1).
The electron quasi-particle excitation across the interface satisfies the
mean-field equations
\begin{subequations}
\label{eq:BdGsimple}
\begin{eqnarray}
\omega \,f\!\left(\bm{r}\right) &=&
-\left[\frac{\nabla^2}{2m}+\frac{k_{\text{F}}^2}{2m}
\right]f\!\left(\bm{r}\right)
+\Delta\!\left(z\right) g\!\left(\bm{r}\right), \\
\omega \,g\!\left(\bm{r}\right) &=&
\left[\frac{\nabla^2}{2m}+\frac{k_{\text{F}}^2}{2m}\right]
g\!\left(\bm{r}\right) + \Delta^*\!\left(z\right) f\!\left(\bm{r}\right),
\end{eqnarray}
\end{subequations}
where the junction has been divided into small 
neighborhoods at positions $\bm{r}$, each being a homogeneous 
system with $k_{\text{F}}$ the Fermi wave vector and $\omega$ with 
$\hbar=1$ the energy measured from the chemical potential, which is in 
the middle of the EI gap because of the conduction-valence band 
symmetry.

The conduction and valence band components of the quasi-particle 
wave function are defined \cite{Andreev}, respectively, by 
$f\!\left(\bm{r},t\right) =
\langle\Psi_0|\tilde{\psi}_b\!\left(\bm{r},t
\right)\!|\Psi_{\bm{k}}^e\rangle$ and
$g\!\left(\bm{r},t\right) = \langle\Psi_0|\tilde{\psi}_a\!\left(\bm{r},t
\right)\!|\Psi_{\bm{k}}^e\rangle$, where
$\tilde{\psi}_b\!\left(\bm{r} \right)$
[$\tilde{\psi}_a\!\left(\bm{r} \right)$] is the annihilation
operator of an electron in the conduction (valence) band
at position $\bm{r}$ and time $t$, and
$|\Psi_{\bm{k}}^e\rangle$ is the excited state, labeled by the index 
$\bm{k}$, obtained by adding
one electron to the exact ground state, $|\Psi_0\rangle$.
The self-consistent order parameter or electron-hole pairing potential
$\Delta\!\left(z\right)$ is a smooth
function of the distance normal to the inferface, $z$, tending 
respectively to the asymptotic values
zero when $z\rightarrow -\infty$, inside the bulk region of SM, and to
the constant $\Delta_0$ when $z\rightarrow +\infty$, inside the bulk
EI [Fig.~2(a)]. We consider the elastic scattering at
equilibrium, matching wave functions of the incident
($I$), transmitted ($C$
and $D$ channels) and reflected ($A$ and $B$) 
states at the boundary (Fig.~1), 
following Refs.~\cite{Andreev,BTK}.
We also calculate the particle density current, $\bm{J}=\rho\bm{v}$, 
where $\bm{v}$
is the semiclassical group velocity and $\rho = \left|f\right|^2
+ \left|g\right|^2$ is the probability density.
\begin{figure}
\centerline{\epsfig{file=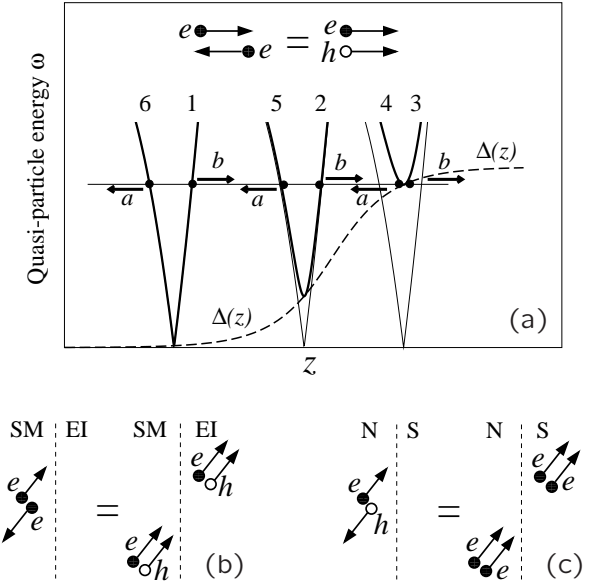,width=3.1in,,angle=0}}
\caption{
(a) ``Andreev'' reflection of an electron of the  conduction band $b$
from the SM/EI interface. 
The thicker (thinner) curves correspond to the renormalized
(``bare'') bands, where the gap $\Delta(z)$ is space dependent.
Only the relevant portion of the spectrum, close to $k_{\text{F}}$,
is shown here.
The numbers 1--6 show the time sequence of the 
reflection from band $b$ to band $a$.
The whole event can be seen 
as a dissipationless flow of electron-hole
pairs (see inset at the top) from the SM to the EI.
(b) Sketch of the elementary scattering process for electrons with
energies within the EI bulk gap $\Delta_0$, reflected on the SM side.
(c) Elementary scattering process at the interface between a normal
metal (N) and a superconductor (S). 
}
\label{fig2}
\end{figure}

There are three qualitatively important results. 
First, {\em all} three cartesian
components of the velocity of the reflected
electron change sign [Fig.~2(b)]. This is not due to 
interface roughness since we take it to be completely flat. The 
exciton condensation causes the dominance of the intravalley 
reflection (from $I$ to $A$ and $C$) but examination of the 
scattering states on the Fermi surfaces in Fig.~1(b) shows 
that the state $A$ is an electron on the valence band Fermi surface 
with complete reversal of the velocity vector which may be regarded 
as a valence hole moving in the opposite direction.
This is reminescent of the Andreev case \cite{Andreev}
when an incident electron is reflected as a hole
from the NS boundary with the velocity reversed [Fig.~2(c)].

Second, the exciton condensate on the EI  side induces exciton order 
on the SM side (the proximity effect). The off-diagonal contribution 
to the electronic density
\protect{$\langle\Psi_0|\tilde{\psi}_b\!\left(\bm{r} \right)
\tilde{\psi}_a^{\dagger}\!\left(\bm{r}\right)
\!|\Psi_0\rangle$} would be zero in
an isolated SM but by the proximity with the condensate
acquires the value
\begin{eqnarray}
&& \langle\Psi_0|\tilde{\psi}_b\!\left(\bm{r} \right)
\tilde{\psi}_a^{\dagger}\!\left(\bm{r}\right)
\!|\Psi_0\rangle  =  \sum_{\bm{k}}f_{\bm{k}}\!\left(\bm{r}\right)
g^*_{\bm{k}}\!\left(\bm{r}\right) \nonumber\\
& \approx & 2\int\!\!\text{d}\,\omega\,{\cal{N}}\!\left(\omega\right)
\cos{\left[\arctan{\left(\frac{\Delta_0}{\omega}\right)}+2
\frac{\omega}{v_{\text{F}}}z\right]},
\label{eq:sum}
\end{eqnarray}
where $v_{\text{F}}=k_{\text{F}}/m$,
and ${\cal{N}}\!\left(\omega\right)$ is the density of states.
Inside the gap ($\omega\approx 0$) each quasiparticle
contributes to the sum (\ref{eq:sum})
with a term $\sim \exp{\left[\text{i}\arctan{\left(\Delta_0/\omega\right)}
+2\text{i}\omega z/v_{\text{F}}\right]}$. The only coordinate dependence
enters this expression via the phase factor, $2\omega z/v_{\text{F}}$,
which represents the relative phase shift of conduction- and
valence-band components of the wavefunction.
If $\omega=0$, then these components keep constant relative phase
$\arctan{\left(\Delta_0/\omega\right)}$
all the way to $z=-\infty$, where no pairing interactions
exist.
Therefore, the reflected electron has exactly the
same velocity as the incident particle,
and will thus retrace {\em exactly the same path} all the way 
to $z=-\infty$. At finite energy, 
the $z$ dependent oscillations provide destructive interference 
on the pair coherence.
Hence, the paths of incident and reflected electrons
part ways away from the interface. Analogous considerations apply to 
the incident electron and to the Andreev-reflected hole in a sub-gap 
scattering event at the NS interface \cite{Zagoskin}.

Third, the ratio of incident electrons
$C(\omega)$ which are transmitted
through the interface depends on the coherence factors
of the condensate and is strongly suppressed close to the gap
(see Appendix).
The dependence of
$C(\omega)$ on $\omega$ and $\Delta_0$ is formally identical to that
for a bogoliubon across a NS interface, provided that the
exciton binding energy $\Delta_0$ is replaced with the BCS gap:
\begin{equation}
C(\omega)=2\sqrt{2\left(\omega-\left|\Delta_0\right|\right)
/\left|\Delta_0\right|} \qquad
\omega\approx\left|\Delta_0\right|.
\label{eq:Cdep}
\end{equation}
Below the gap the electron is totally ``Andreev'' reflected and
the transmission is zero.

\begin{figure}
\centerline{\epsfig{file=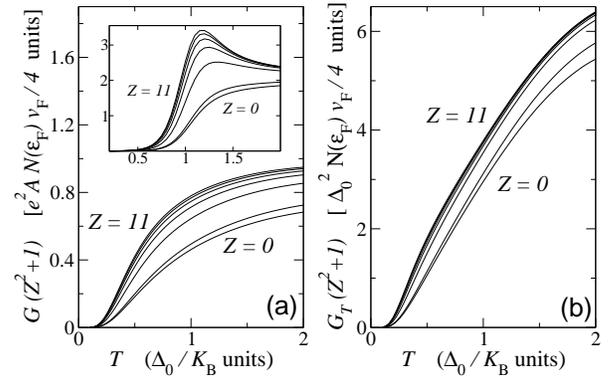,width=3.1in,,angle=0}}
\caption{
(a) Interface conductance $G$ vs.~$T$ for different
values of the dimensionless strength of the barrier at the interface,
\protect{$Z=0,1,3,5,7,9,11$}.
The shown values have been divided by the constant transmission coefficient
when $\Delta_0=0$,
$C_{\Delta_0=0}=(Z^2+1)^{-1}$. 
Inset: Differential conductance [units of 
$e^2{\cal{A}}N\!(\varepsilon_{\text{F}})\,v_{\text{F}}/4$] 
vs.~$V$ (units of $\Delta_0/e$), at 
$K_{\text{B}}T/\Delta_0=0.1$.
(b) Plot for the thermal conductance $G_T$, analogous to 
(a).
}
\label{figGT}
\end{figure}
From the 
results for transmission and
reflection probabilities, 
we derive in the linear response regime the values of the electrical and
thermal interface conductances, $G$ and $G_T$,
respectively \cite{us,usunpublished}. The Seebeck coefficient is 
zero due to the symmetry artifact of the model \cite{Ziman2}.
Then, except for an additive phonon contribution to the thermal conductance,
the interface thermoelectric properties are completely determined by
$G$ and $G_T$. 
The $Z=0$ curves in Fig.~3 show
that $G$ and $G_T$ have an activation threshold at low temperature, 
$T$, proportional
to the gap $\Delta_0$. If $K_{\text{B}}T\ll \left|\Delta_0\right|$, 
$G_T$ has exactly the same
functional dependence on $T$ and $\Delta_0$
in the SM/EI as in the NS junction.
Remarkably, the rate of entropy production \cite{Ziman2},
$\dot{S} = G (\delta V)^2/T + {\cal A} G_T (\delta T)^2/T^2 $,
where $\delta T$ and $\delta V$ are the temperature and voltage drops
at the interface and ${\cal A}$ the cross-sectional area,
is the same very low value in both cases.
In the NS junction the term proportional to $G$, unrelated
to the superfluid component, comes from bogoliubons which,
when they cross the NS interface, experience
the same resistance $G^{-1}$ as electrons do across the SM/EI boundary.

To shed light on the dissipationless motion of electrons 
in the linear transport regime,
we place a $\delta$-function potential barrier at $z=0$,
$H\delta(z)$, simulating the effect of including 
a thin insulating layer \cite{BTK}. 
Coherence between two sides of the interface
is diminished as the dimensionless barrier strength,
$Z = mH/k_{\text{F}}$, increases from zero (clean junction)
to finite values (tunneling regime). 
Figure 3 displays $G$ and $G_T$
for increasing values of $Z$. Since the transmission coefficient $C(\omega)$
decreases uniformly in the absence
of any electron-hole pairing (see Appendix), 
$C_{\Delta_0=0}=(Z^2+1)^{-1}$,
we rescale conductances dividing them by $C_{\Delta_0=0}$.
Naively, we would expect that the insertion of an insulating layer
would reduce the conductances.
On the contrary, the effect is just the opposite: as $Z$ increases,
$G/C_{\Delta_0=0}$ and $G_T/C_{\Delta_0=0}$ {\em increase}, eventually
reaching saturation in the tunneling regime. 
This shows that the exciton order induced in the SM side by EI 
makes the junction less conductive for charge and heat transport.
The plot of the differential conductance 
vs.~$V$ at low $T$ [inset of Fig.~3(a)] 
allows clear monitoring of the transition from the transparent to the 
opaque limit, where transport is recovered. 
The effect is maximum for $eV\approx \Delta_0$ and as $T\rightarrow 0$,
when the diff.~conductance becomes proportional to  
$C(eV)+D(eV)$.

These features which distinguish the exciton insulator from the normal 
insulating state may be explained by two alternate physical pictures. 
The conventional view is that electrons below the energy gap cannot 
contribute to transport as they are back scattered by the gap 
barrier, $\Delta(\bm{r})$, formed by proximity effect of the 
EI. The less conventional view is to make use of the 
analogy with the NS junction. Instead of 
counting the electrons in the valence band as negatively charged 
carriers of the current, we may start with the state with the valence 
band filled to the top as carrying zero current even under an 
electrical or thermal current and regard each unoccupied state in the 
valence band as a positively charged carrier --- a hole --- moving in the 
direction opposite to the electron. Then the reflected electrons are 
replaced by incoming holes towards the barrier.
Therefore, the incident conduction electron and the valence hole may 
be viewed as a correlated pair moving  towards the interface 
[Fig.~2(b)].
The novelty is that a constant
electron-hole current moves from the SM to the EI below the gap, where
electric transport is blocked.
As the electron-hole pair approaches the interface
from the SM side, the {\em exciton} current
is converted into the condensate {\em supercurrent}:
the global effect is that in the steady state an exciton current exists
flowing constantly and reversibly all the way from the SM to the EI
without any form of dissipation.

The above scenario follows from the continuity equation
for the electron-hole current. The probability density
$\rho_{\text{e-h}}\!\left(\bm{r},t\right)$ for finding either 
a conduction-band
electron or a valence-band hole at a particular time and place is
\protect{$\rho_{\text{e-h}}\!\left(\bm{r},t\right)=
\left|f\right|^2 + 1 - \left|g\right|^2$}.
Thus, the associated continuity equation is
\begin{equation}
\frac{\partial \rho_{\text{e-h}}}{\partial t}+\nabla\cdot
{\bm{J}_{\text{e-h}}}=0, \qquad
\bm{J}_{\text{e-h}} = \bm{J}_{\text{pair}} + \bm{J}_{\text{cond}}.
\label{eq:continuity2}
\end{equation}
One component of the electron-hole current, 
\protect{$\bm{J}_{\text{pair}}=m^{-1}{\text{Im}}\{
f^*\nabla f + g^*\nabla g\}$}, is the
density current of the electron-hole pair, similar to 
the standard particle carrier \protect{$\bm{J}=m^{-1}{\text{Im}}\{
f^*\nabla f - g^*\nabla g\}$} with an important difference in sign.
The other component,  \protect{$\nabla\cdot {\bm{J}_{\text{cond}}} =
-4\,{\text{Im}}\{ f^* g \, \Delta \}$},
depends explicitly on the built-in coherence of the electron-hole 
condensate $\Delta$, and may be described as the exciton 
supercurrent of the EI state.

Going back to our picture of $\Delta\!\left(z\right)$ smoothly varying
in space (Fig.~2), if $\omega < \left|\Delta_0\right|$,
each electron wave function, solution of Eq.~(\ref{eq:BdGsimple}),
carries zero total electric current $e\bm{J}$, which is the
sum of the equal and opposite
incident and reflected fluxes, and finite and constant
electron-hole current \protect{$\bm{J}_{\text{e-h}}=2v_{\text{F}}\bm{n}$},
with $\bm{n}$ a unit vector.
When $z\rightarrow -\infty$, far from the interface on the SM side,
the supercurrent contribution $\bm{J}_{\text{cond}}$
is zero.
As $z$ increases and $\Delta\!\left(z\right)$ gradually rises,
both $\bm{J}$ and \protect{$\bm{J}_{\text{e-h}}$} conserve
their constant value, independent of $z$, since quasi-particle states
are stationary. However, their analysis in terms of incident and
reflected quasi-particles is qualitatively different.
From the electron point of view, we see in Fig.~2(a) that
the incoming conduction band particle approaching the EI boundary
sees its group velocity progressively reduced, up to the classical
turning point where it changes direction and branch of the spectrum:
there is no net electric current.
From the exciton point of view,
as the contribution to the electron-hole current $\bm{J}_{\text{pair}}$
vanishes approaching the boundary, since the group velocity
goes to zero at the classical turning point
where the wavefunction becomes evanescent, $\bm{J}_{\text{pair}}$ is
converted into the supercurrent $\bm{J}_{\text{cond}}$.
Excitons therefore can flow into the EI side without any resistance,
and the sum \protect{$\bm{J}_{\text{e-h}}$} of the two contributions,
$\bm{J}_{\text{pair}}$ and $\bm{J}_{\text{cond}}$,
is constant through all the space (Fig.~2).
As $\omega$ exceeds $\left|\Delta_0\right|$, $\bm{J}$ acquires a finite
value and \protect{$\bm{J}_{\text{e-h}}$} monotonously decreases. However,
close to the gap, electron transmission to the EI side is
still inhibited [cf.~Eq.~(\ref{eq:Cdep})] by the pairing between
electrons and holes of the condensate: an electron can stand alone and
carry current only after its parent exciton has been ``ionized''
by injecting --- say ---  a conduction-band electron or by filling
a valence-band hole in the EI. The ionization costs
an amount of energy of the order of the binding energy of the
exciton, $\left|\Delta_0\right|$.
Therefore, as long as $\omega \approx \left|\Delta_0\right|$,
the competition between exciton and electron flow favors 
intravalley
reflection, which is the source of both electric and thermal resistances.
In equilibrium, there is no net charge or heat
flow, since quasi-particles with $\bm{v}$ and $-\bm{v}$
compensate each other. However, if a heat current flows,
the net drift velocity of electrons and holes locally
``drags'' the exciton supercurrent, which otherwise would be pinned
by various scattering sources \cite{Keldyshmaligno}.
 
Real-system candidates for the experimental study of the
thermoelectric properties of the SM/EI junction include
a few rare-earth calcogenides such as TmSe$_x$Te$_{1-x}$ \cite{exp}.
Alternatively, transport in
the direction perpendicular to planes of layered graphite
could reveal a latent excitonic insulator 
instability \cite{Khveshchenko}. 
Also, a lateral junction made of coupled quantum
wells where conduction and valence bands are spatially separated
appears very promising \cite{Datta,butov}.
The latter system would provide a convenient setup
to compare homo- and hetero-junctions, since inter-band hybridization
could be easily controlled by inter-layer tunneling.

We thank (M.R.) E.~Randon, E.~K.~Chang, C.~Tejedor,
(M.R. and L.J.S.) L.~V.~Butov, J.~E.~Hirsch,
for stimulating discussions, and M.~Fogler for bringing to our attention 
the related papers on a junction between metal and Peierls charge
density wave \cite{fogler}.
This work is supported (L.J.S.) by NSF DMR 0099572 and 0403465.

\appendix

\section{Summary}

This Appendix is organized as follows:
After a description of the model
and of the equations, we present the solution 
of the electron motion across the
interface (App.~\ref{model}), we 
constrast the homogeneous junction with the
heterogeneous junction (App.~\ref{s:scattering}),
and then we explore in detail the transport through
the homogeneous junction (App.~\ref{Andreevm}).

\section{The two-band model}\label{model}

\subsection{The junction}

We describe the junction that a semimetal (SM) forms with a
small-gap semiconductor (SC), either a material where the gap comes
from band hybridization (NSC), case (i), or an excitonic insulator
(EI), case (ii),
by means of a spinless two-band Hamiltonian
\begin{equation}
\mathcal{H}=\mathcal{H}_{0}+\mathcal{H}_{1}+\mathcal{H}_{2}.
\label{eq:H}
\end{equation}
Here $\mathcal{H}_{1}$ is the kinetic term which embodies the effect
of the ideal and frozen crystal lattice on electrons,
in terms of the envelope function in the effective mass approximation:
\begin{equation}
\mathcal{H}_{1}=\sum_{i=a,b}\int \!\!\text{d}\,\bm{r}\, \psi_i^{\dagger}
\!\left(\bm{r}\right)\varepsilon_i\!\left(\bm{r}\right)
\psi_i\!\left(\bm{r}\right).
\label{eq:h1}
\end{equation}
The field operator $\psi_a\!\left(\bm{r}\right)$
[$\psi_b\!\left(\bm{r}\right)$] annihilates an electron in the valence
(conduction) energy band at the position $\bm{r}$ in space.
We assume that the two bands are isotropic in the 
wave-vector space: the valence-band has
a single maximum at $\bm{k}=0$ while the conduction-band a single minimum at
$\bm{k}=\bm{w}$. We ignore complications due to the
presence of equivalent minima.
We write the single-particle energies as
\begin{subequations}
\label{eq:bandstructure}
\begin{eqnarray}
\varepsilon_a\!\left(\bm{k}_a\right)&=&-G/2-(2m_a)^{-1}k_a^2,\\
\varepsilon_b\!\left(\bm{k}_b\right)&=&G/2+(2m_b)^{-1}k_b^2,
\end{eqnarray}
\end{subequations}
where $\bm{k}_a$ and $\bm{k}_b$ refer to the respective band extrema,
and $m_a$ and $m_b$ are (positive) effective masses \cite{Kohn}.
Throughout this work we put $\hbar=1$ and assume that the system has
unit volume, unless explicitly stated otherwise.  Energies are 
measured from the center of the gap $G$ which may be positive or negative.
Note that the chemical potential $\mu$ is not necessarily zero.
We assume that the total number of electrons in the two bands
is such that for positive $G$ we have an intrinsic semiconductor,
namely there is one spinless electron per unit cell.
For negative $G$ we have, in the absence of interactions or
other potentials (\protect{$\mathcal{H}_{0}=\mathcal{H}_{2}=0$}),
a semimetal with Fermi wave vector given by
\begin{equation}
k_{\text{F}}^2=-2m_{\text{red}} G,
\label{eq:kf}
\end{equation}
where $m_{\text{red}}$ is the reduced mass
\begin{equation}
m_{\text{red}}^{-1}=m_a^{-1}+m_b^{-1}.
\end{equation}
According to Eq.~(\ref{eq:kf}), $k_{\text{F}}$ is imaginary for
a semiconductor ($G>0$).
The band operators in real space $\varepsilon_i\!\left(\bm{r}\right)$
appearing in Eq.~(\ref{eq:h1}) are defined as follows:
\begin{subequations}
\begin{eqnarray}
\varepsilon_a\!\left(\bm{r}\right) &=& -G/2+(2m_a)^{-1}\nabla^2;\\
\varepsilon_b\!\left(\bm{r}\right) &=& G/2-(2m_b)^{-1}\nabla^2.
\end{eqnarray}
\end{subequations}

The two-body term $\mathcal{H}_{2}$ describes
the inter-band Coulomb interaction
\begin{eqnarray}
\mathcal{H}_{2} &=&
\int \!\!\text{d}\,\bm{r}\,\text{d}\,\bm{r'}\,
\psi_a^{\dagger}\!\left(\bm{r}\right)
\psi_b^{\dagger}\!\left(\bm{r'}\right) \nonumber\\
&& \times \quad V_2\!\left(\bm{r-r'}\right)
\psi_b\!\left(\bm{r'}\right)\psi_a\!\left(\bm{r}\right),
\end{eqnarray}
where $V_2\!\left(\bm{r}\right)$ is the dielectrically screened Coulomb
potential \cite{Keldysh}. The two sides of the homogeneous junction
are described by the variation in space
of the mean-field electron-hole pairing potential $\Delta$ across the
boundary plane between the two emispaces [see Fig.~2(a)].
Well inside the SM $\Delta$ is zero, then it smoothly rises through
the interface and eventually takes a constant value in the bulk of the EI.
Therefore, the two sides of the homogeneous
junction only differ in the value of
the pairing potential $\Delta$, in close analogy to the situation
for the intermediate state between the normal and superconducting
phases of a metal. For the heterogeneous junction, $\Delta=0$ and
the discontinuity is brought about by the hybridization
term we introduce below.

The one-body term $\mathcal{H}_{0}$ is the sum of two parts,
\begin{equation}
\mathcal{H}_{0}=\mathcal{V} + \mathcal{V}_{\text{hyb}}.
\end{equation}
$\mathcal{V}$ is the intra-band term,
\begin{equation}
\mathcal{V}=\sum_{i=a,b}\int \!\!\text{d}\,\bm{r}\,
\psi_i^{\dagger}\!\left(\bm{r}\right)V\!\left(\bm{r}\right)
\psi_i\!\left(\bm{r}\right),
\end{equation}
which takes into account, via the single-particle potential
$V\!\left(\bm{r}\right)$, the band offset or
the possible impurities and defects at the
junction interface, such as a thin insulating layer.
The potential $V\!\left(\bm{r}\right)$ can also describe the effect of
a voltage bias applied to the junction in a steady-state regime.
In this latter case, a rigorous approach would
require $V\!\left(\bm{r}\right)$ to be determined in a self-consistent way
together with the electronic charge distribution.
The inter-band term,
\begin{equation}
\mathcal{V}_{\text{hyb}}=\int \!\!\text{d}\,\bm{r}\,
\psi_b^{\dagger}\!\left(\bm{r}\right)V_{\text{hyb}}\!\left(\bm{r}
\right) \psi_a\!\left(\bm{r}\right)\quad + \text{\;\,H.c.},
\end{equation}
describes the hybridization of conduction and valence bands by means of
the potential 
$V_{\text{hyb}}\!\left(\bm{r}\right)$. Its symmetry
properties depend on the characters of $a$ and $b$ bands. For a discussion
of the influence of such term on exciton condensation see
Refs.~\cite{ferro,ferro2,lu}.
For the homogeneous junction $V_{\text{hyb}}=0$.

Renormalization effects due to
intra-band Coulomb interaction and temperature dependence are already
included in the energy band structure (\ref{eq:bandstructure}),
whose parameters are assumed to be known.

\subsection{The equations of motion}

We follow Andreev \cite{Andreev} and introduce the electron
quasi-particle amplitudes
\begin{subequations}
\begin{eqnarray}
f\!\left(\bm{r},t\right) &=& \langle\Psi_0|\tilde{\psi}_b\!\left(\bm{r},t
\right)\!|\Psi_{\bm{k}}^e\rangle,\\
g\!\left(\bm{r},t\right) &=& \langle\Psi_0|\tilde{\psi}_a\!\left(\bm{r},t
\right)\!|\Psi_{\bm{k}}^e\rangle.
\end{eqnarray}
\label{eq:fgdef}
\end{subequations}
Here $|\Psi_0\rangle$ and $|\Psi_{\bm{k}}^e\rangle$ are respectively
the true interacting ground state and the excited state with one
electron added to the system; the latter state is labeled by the
quantum index $\bm{k}$, which not should be necessarily identified
with the crystal momentum, the translational symmetry being destroyed
by the presence of the junction. States and operators are written
in the Heisenberg representation \cite{AGD}:
\begin{eqnarray}
\tilde{\psi}_i\!\left(\bm{r},t\right) &=&
\exp\!\left(\text{i}\left[\mathcal{H}
-\mu N\right]t\right) \nonumber\\
&& \times \quad \psi_i\!\left(\bm{r}\right)
\exp\!\left(-\text{i}\left[\mathcal{H}
-\mu N\right]t\right).
\label{eq:defop}
\end{eqnarray}
Since we also consider non-equilibrium situations,
the chemical potential $\mu\!\left(\bm{r}\right)$ can vary in space.
The number operator $N$ is defined by
\begin{equation}
\mu N=\sum_{i=a,b}\int \!\!\text{d}\,\bm{r}\,  \mu(\bm{r}) \,
\psi_i^{\dagger} \!\left(\bm{r}\right) \psi_i\!\left(\bm{r}\right).
\label{eq:numop}
\end{equation}
Writing down the Heisenberg equations of motion for the operators
$\tilde{\psi}_i\!\left(\bm{r},t\right)$ and simplifying them by means
of the mean-field approximation, we derive a set of two coupled
integro-differential equations for the amplitudes
$f\!\left(\bm{r},t\right)$ and $g\!\left(\bm{r},t\right)$:
\begin{subequations}
\label{eq:BdG}
\begin{eqnarray}
&&\text{i}\frac{\partial f\!\left(\bm{r},t\right)}{\partial t} =
\left[\varepsilon_b\!\left(\bm{r}\right) + V\!\left(\bm{r}\right)
-\mu\!\left(\bm{r}\right)\right]f\!\left(\bm{r},t\right) \nonumber\\
&+&\!\!\!\int \!\!\!\text{d}\bm{r'}
\left[\Delta\!\left(\bm{r},\bm{r'}\right)\!
+\!\delta(\bm{r}-\bm{r'})V_{\text{hyb}}\!\left(\bm{r}\right)\right]
\!g\!\left(\bm{r'},t\right),\label{eq:BdGa} \\
&& \text{i}\frac{\partial g\!\left(\bm{r},t\right)}{\partial t} =
\left[\varepsilon_a\!\left(\bm{r}\right) + V\!\left(\bm{r}\right)
-\mu\!\left(\bm{r}\right)\right]g\!\left(\bm{r},t\right) \nonumber\\
&+&\!\!\!\int \!\!\!\text{d}\bm{r'}
\left[\Delta^*\!\left(\bm{r'},\bm{r}\right)\!
+\!\delta(\bm{r'}-\bm{r})V_{\text{hyb}}^*\!\left(\bm{r}\right)\right]
\!f\!\left(\bm{r'},t\right).\label{eq:BdGb}
\end{eqnarray}
\end{subequations}
The built-in coherence of the exciton condensate,
$\Delta\!\left(\bm{r},\bm{r'}\right)$, appearing in Eqs.~(\ref{eq:BdG})
is determined self-consistently from
\begin{equation}
\Delta\!\left(\bm{r},\bm{r'}\right)=V_2\!\left(\bm{r-r'}\right)
\langle\Psi_0|\tilde{\psi}_b\!\left(\bm{r}\right)
\tilde{\psi}_a^{\dagger}\!\left(\bm{r'}\right)|\Psi_0\rangle.
\label{eq:delta}
\end{equation}
Apart from the factor $V_2\!\left(\bm{r-r'}\right)$,
$\Delta\!\left(\bm{r},\bm{r'}\right)$ defined in Eq.~(\ref{eq:delta})
represents the wavefunction of the electron-hole condensate,
smoothly vanishing when $\left|\bm{r-r'}\right|$ is larger than the
characteristic exciton radius.
 From the value of $\Delta\!\left(\bm{r},\bm{r}\right)$ we identify the
different regions of the homogeneous junction, when $V_{\text{hyb}}=0$:
$\Delta=0$ refers to the bulk SM, $\Delta$ maximum
to the bulk EI, while the
value of $\Delta$ in the interface region changes in a slow and
continuous manner from one bulk limit to the other [see Fig.~2(a)].

The amplitudes $f\!\left(\bm{r},t\right)$ and $g\!\left(\bm{r},t\right)$
are the position space representation of the stationary electron-like
elementary excitation across the {\em whole} junction. Taken
together, they signify the wave function of the quasi-particle: $f$ ($g$)
is the component of the probability amplitude for an electron
of belonging to the conduction (valence) band. They satisfy the
normalization condition
\begin{equation}
\int \!\!\text{d}\,\bm{r}\,\left[\left|f\!\left(\bm{r},t\right)\right|^2
+\left|g\!\left(\bm{r},t\right)\right|^2\right]=1,
\end{equation}
and have always positive excitation energy $\omega$ due to the
definitions (\ref{eq:fgdef}-\ref{eq:defop}).
The hole-like excitation amplitudes are
analogous to (\ref{eq:fgdef}) and satisfy two coupled 
equations that are the complex conjugates of Eqs.~(\ref{eq:BdG}).
The probability current density
$\bm{J}\!\left(\bm{r},t\right)$ can be found starting from the
probability density
$\rho\!\left(\bm{r},t\right)$ for finding either a conduction-
or a valence-band electron at a particular time and place
\cite{notecurrent}, defined as
\protect{$\rho\!\left(\bm{r},t\right)=
\left|f\right|^2 +\left|g\right|^2$}. After some manipulation of
the equations of motion (\ref{eq:BdG}), one derives the continuity
equation
\begin{equation}
\frac{\partial \rho}{\partial t}+\nabla\cdot {\bm{J}}=0,
\label{eq:continuity}
\end{equation}
where
\begin{equation}
\bm{J}={\text{Im}}\!\left\{
f^*\frac{\nabla}{m_b} f - g^*\frac{\nabla}{m_a} g\right\}.
\label{eq:J}
\end{equation}
Note that the two terms appearing in the rhs of Eq.~(\ref{eq:J}), referring
respectively to conduction and valence band electrons, have
opposite sign since the curvature of the two bands is opposite.
One can verify that the semiclassical group velocity of the quasiparticle,
\protect{$\bm{v}_{\text{g}}=\nabla_{\bm{k}}\, \omega $},
coincides with the velocity
$\bm{v}$ given by the full quantum mechanical expression (\ref{eq:J}),
with \protect{$\bm{J}=\rho\,\bm{v}$}.

The SC bulk solution of Eqs.~(\ref{eq:BdG}) [with $V\!\left(\bm{r}\right)
=0$ and $\mu$ constant] is given by conduction- and valence-band
plane waves coupled together,
\begin{equation}
{f_{\bm{k}}\!\left(\bm{r},t\right) \choose
g_{\bm{k}}\!\left(\bm{r},t\right)} =
{ u_{\bm{k}} \choose v_{\bm{k}} } {\rm e}^{ \text{i}\left(\bm{k\cdot r}
-\omega t \right) },
\label{eq:bulk}
\end{equation}
with energy
\begin{equation}
\omega\!\left(\bm{k}\right)=
\frac{\tilde{\varepsilon}_b\!\left(\bm{k}\right) +
\tilde{\varepsilon}_a\!\left(\bm{k}\right) }{2}  +
\sqrt{\xi_{\bm{k}}^2+\left|V_{\text{hyb}}(\bm{k})
+\Delta_{\bm{k}}\right|^2},
\label{eq:dispersiong}
\end{equation}
where \protect{$\tilde{\varepsilon}_i\!\left(\bm{k}\right)=
\varepsilon_i\!\left(\bm{k}\right)-\mu$,}
\protect{$\xi_{\bm{k}}=\left[\varepsilon_b\!\left(\bm{k}\right)
-\varepsilon_a\!\left(\bm{k}\right)\right]/2$,}
$\Delta_{\bm{k}}$ is the Fourier component of $\Delta\!\left(\bm{r}\right)$,
and similarly $V_{\text{hyb}}(\bm{k})$.
Here we exploit the fact that
$\Delta\!\left(\bm{r},\bm{r'}\right)=\Delta\!\left(\bm{r-r'}\right)$,
due to translational invariance of the bulk.
The amplitudes are such that
\begin{equation}
\left|u_{\bm{k}}\right|^2=\frac{1}{2}\!
\left(1+\frac{\xi_{\bm{k}}}{E_{\bm{k}}}
\right),\qquad \left|u_{\bm{k}}\right|^2+\left|v_{\bm{k}}\right|^2=1,
\end{equation}
and the relative phase between $u_{\bm{k}}$ and $v_{\bm{k}}$ is given by
\begin{equation}
\frac{u_{\bm{k}}}{v_{\bm{k}}}=
\frac{\Delta_{\bm{k}}+V_{\text{hyb}}(\bm{k})}{E_{\bm{k}}-\xi_{\bm{k}}}.
\end{equation}
Here \protect{$E_{\bm{k}}=\sqrt{\xi_{\bm{k}}^2+\left|\Delta_{\bm{k}}
+V_{\text{hyb}}(\bm{k}) \right|^2}$}.
When $\Delta=0$, the amplitude (\ref{eq:bulk}) is
always an admissible solution. However, if excitons condense,
(\ref{eq:bulk}) is sustainable
only if the self-consistency condition derived by the definition
of $\Delta$ is satisfied. This condition, which
can be easily obtained from Eq.~(\ref{eq:delta}), is formally analogous
to the BCS gap equation
\begin{equation}
\Delta_{\bm{k}}=\sum_{\bm{p}}
\frac{V_{2,\,\bm{k-p}}
\left[\Delta_{\bm{p}}+V_{\text{hyb}}(\bm{p})\right]}{2E_{\bm{p}}}.
\label{eq:gap}
\end{equation}
Here $V_{2,\,\bm{k}}$ is the Fourier component of
$V_2\!\left(\bm{r}\right)$.
A sufficiently strong hybridization can destroy the pairing 
\cite{ferro,ferro2}.
Either the hybridization of conduction and valence band
or the pairing $\Delta_{\bm{k}}$
open a gap in the excitation energy bands. This gap is
suppressed in the ``normal'' SM state, where $V_{\text{hyb}}=0$ and
the gap equation (\ref{eq:gap}) admits only
the trivial solution $\Delta_{\bm{k}}=0$. In this latter case
the quasi-particles are simply conduction-
or valence-band plane waves.

There is a suggestive parallelism between the
formalism used here and the treatment of elementary single-particle
excitations in conventional superconductors.
However, while bogoliubons in the
BCS theory have a mixed electron/hole character, here the elementary
excitations carry an integer electron charge. In addition, the
momentum $\bm{k}$ carried by the quasi-particle now has a mixed character,
since the solution (\ref{eq:bulk}) couples momenta of different displaced
energy bands. This momentum mixing reflects the new periodicity in
real space of the ``distorted'' broken-symmetry ground state,
as can be seen by the additional long-range order shown by the
electron density, characterized by the wave vector
$\bm{w}$ \cite{Kohn,KohnRMP}.

\section{Scattering at the interface}\label{s:scattering}


The interface potential is physically different for the two types of
dielectrics.
In the case of the ordinary hybridized-band SC, the interface potential is
always sharp and discontinuous, no matter the quality of sample
preparation and processing. The reason is that, in this
{\em heterogeneous} junction --- the usual case ---,
the two sides correspond to different bulk regions with their own
different band structures. Moreover, the energy bands in the interface
region can be displaced by any local band-offset potential,
and growth-dependent impurities, surface states, oxidation layers can act
as potential barriers or traps. The {\em homogeneous} junction,
instead, corresponds to the {\em same} material undergoing a spatially
dependent exciton condensation. Even if the junction
is made of two parts having a slightly different chemical composition,
it is not possible to identify
a sharp interface because, due to the proximity effect discussed
in the main text, the pairing potential $\Delta\!\left(\bm{r}\right)$
naturally extends beyond the original EI side and gently vanishes
as it enters the SM side [Fig.~2(a)].
This type of junction can be realized by applying
a gradient pressure to an EI sample or by spatially-varying doping, exploiting
the fact that impurities destroy the electron-hole coherence \cite{Zittartz}.

In the two-band picture, there are two possible channels for
below-gap reflection of carriers coming from the SM side:
intra-band ($B$ channel in Fig.~1) and inter-band ($A$ channel).
Which channel dominates ultimately depends on
the nature of the interface potential,
as Sham and Nakayama discussed in the context of the MOSFET
space-charge layer \cite{SNter}. We find that the $A$ ($B$)
channel characterizes the scattering at the homogeneous
(heterogeneous) interface.

The simplest case we investigate systematically is the
one-dimensional system wth conduction and valence bands having the
same curvature, namely $m_b=m_a=m$ (Fig.~1).
The formal treatment remarkably resembles that of the metal-superconductor
junction, as e.g.~in the theory of Blonder,
Tinkham and Klapwijk (BTK) \cite{BTK}.
We consider the dynamical equilibrium
with no applied voltage across the junction: in this case $\mu(z)=0$
everywhere due to symmetry, where $z$ is the coordinate
orthogonal to the interface plane, which is located at $z=0$.
We describe in a simple way both types of junction
by introducing the generic parameter $X$: for the heterogeneous
junction $X$ coincides with the hybridization potential and
the pairing potential is zero, i.e.~\protect{$X\!\left(z,z'\right)=
\delta(z-z')V_{\text{hyb}}\!\left(z\right)$} and $\Delta\!\left(z,
z'\right)=0$, while we take the opposite for the homogeneous junction,
namely \protect{$X\!\left(z,z'\right)=
\Delta\!\left(z,z'\right)$} and $V_{\text{hyb}}\!\left(z\right)=0$.
We also assume that $X$ is zero for $z<0$, namely $X\!\left(z,
z'\right)=0$ if $z<0$ or $z'<0$ (SM side),
and that it takes a costant value for $z>0$, i.e.~$X\!\left(z,
z'\right)=X\,\delta\!(z-z')$ (gapped SC).
The difference between homogeneous and heterogeneous junction originates
from the range of values allowed for $X$ in the two cases and by the possible
occurrence of an additional interface potential, $V(z)$.

Although, in general, both the built-in coherence $\Delta$
and the hybridization potential $V_{\text{hyb}}$ are complex,
for simplicity we take them to be real and positive.
For the case treated here, this involves no loss of generality: phase
possibly plays a role only in the EI/EI junction.
Under these simplifications, the generic bulk SC excitation energy assumes
the form
\begin{equation}
\omega(k)=E_k=\sqrt{\left(k^2/2m-k^2_{\text F}/2m
\right)^2+X^2},
\label{eq:dispersion}
\end{equation}
which is identical to the BCS dispersion relation for bogoliubons.
Equation (\ref{eq:dispersion}) is equivalent to
the general dispersion relation (\ref{eq:dispersiong}) for
equal masses except it has $X$, independent on $\bm{k}$,
replacing \protect{$V_{\text{hyb}}(\bm{k})+\Delta_{\bm{k}}$.
The latter assumption implies a
long-wavelength regime which holds as long as: (i) we consider
low-energy excitations close to the
bottom of the quasi-particle energy band (ii) assume that
the wavefunction $\Delta$ for the relative motion between electrons
and holes of the condensate varies slowly in space (iii)
take the hybridization term $\bm{k}$-independent and $s$-symmetrical.

In the elastic scattering process at equilibrium,
all relevant quasi-particle
states are those degenerate --- with energy $\omega$ --- on both sides
of the junction (e.g., in Fig.~1(a) the four states with wave vector
$\pm q^+$, $\pm q^-$ on the SM side and the other four with
$\pm k^+$, $\pm k^-$ on the SC side).
We handle the interface by matching wave functions
of the incident, transmitted, and reflected particles at the
boundary. The computation of the fraction of
current carried by the different fluxes allows us to compute
the transport characteristics of the junction.

\subsection{Wavefunction matching}

In the bulk SC, since only the square of $k$ enters the expression
(\ref{eq:dispersion}) for $E_k$,
there will be a pair of magnitudes of $k$ associated with $\omega$,
namely,
\begin{equation}
k^{\pm}=\sqrt{2m}\sqrt{ k^2_{\text F}/2m \pm
\left(\omega^2 - X^2\right)^{1/2} }.
\label{eq:kvect}
\end{equation}
The total degeneracy of relevant states for each $\omega$ is fourfold:
$\pm k^{\pm}$, as sketched in Fig.~1(a).
The two states $\pm k^+$
have a dominant conduction-band character, while the two states $\pm k^-$
are mainly valence-band states. We use the
analytic continuation of Eq.~(\ref{eq:kvect}) when $\omega<X$
or $k^2_{\text F}/2m < \left(\omega^2 - X^2\right)^{1/2}$ and
$k$ is complex,
choosing the Riemann sheet for which wavefunctions are evanescent
and tend to zero as $z\rightarrow \infty$.
Using the notation
\begin{equation}
\Psi(z)=
{f\!\left(z\right) \choose
g\!\left(z\right)}
\end{equation}
the wave functions degenerate in $\omega$ are
\begin{equation}
\Psi_{\pm k^+}=
{ u_0 \choose v_0 } {\rm e}^{ \pm\text{i} k^+z }
\qquad
\Psi_{\pm k^-}=
{ v_0 \choose u_0 } {\rm e}^{ \pm\text{i} k^-z },
\label{eq:bulk1D}
\end{equation}
with the amplitudes $u_0,v_0$ defined as
\begin{eqnarray}
u_0&=&\sqrt{\frac{1}{2}\left[1+\frac{(\omega^2-X^2)^{1/2}}{\omega}
\right]},\nonumber\\
v_0&=&\sqrt{\frac{1}{2}\left[1-\frac{(\omega^2-X^2)^{1/2}}{\omega}
\right]},
\end{eqnarray}
possibly extended in the complex manifold.
With regards to the SM bulk,
$X=0$ and the two possible magnitudes of the momentum $q$ reduce to
\begin{equation}
q^{\pm}=\sqrt{2m}\sqrt{ k^2_{\text F}/2m \pm (\omega -H_{\text{offset}}) }
\label{eq:qvect}
\end{equation}
with wave functions
\begin{equation}
\Psi_{\pm q^+}=
{ 1 \choose 0 } {\rm e}^{ \pm\text{i} q^+z }
\qquad
\Psi_{\pm q^-}=
{ 0 \choose 1 } {\rm e}^{ \pm\text{i} q^-z }
\label{eq:SMbulk1D}
\end{equation}
for conduction and valence bands, respectively
[Fig.~1(a)]. Here we have
introduced for $z<0$ an additional constant
external potential, $V(z)=H_{\text{offset}}$,
which takes into account the band offset at the interface.
The reference chemical potential is zero and lies in the
middle of the SC energy gap $2X$.

Let us work out the boundary conditions on these steady-state plane-wave
solutions at the interface.
The elastic scattering wich occurs at the interface, due e.g.~to the
effect of an oxide layer in a point contact or the localized disorder
in the neck of a short microbridge, is modeled by a
$\delta$-function potential, namely $V\!(z)=H\delta\!(z)$.
The appropriate boundary conditions, for particles traveling from
SM to SC are as follows: (i) Continuity of
$\Psi$ at $z=0$, so \protect{$\Psi_{\text{SC}}(0)=\Psi_{\text{SM}}(0)
\equiv \Psi(0)$}. (ii) \protect{$\left[f_{\text{SC}}'(0)-
f_{\text{SM}}'(0)\right]/(2m)=Hf(0)$} and
\protect{$\left[g_{\text{SC}}'(0)-
g_{\text{SM}}'(0)\right]/(2m)=-Hg(0)$}, the derivative boundary
conditions appropriate for $\delta$-functions
\cite{Boundary}. (iii) Incoming (incident),
reflected and transmitted wave directions are defined by their group
velocities. We assume the incoming conduction band electron produces
only outgoing particles, namely an electron incident from the left can
only produce transmitted particles with positive group velocities
$v_{\text{g}}>0$
and reflected ones with $v_{\text{g}}<0$.
The self-consistency of the equilibrium state
provides a useful check on the computed probability coefficients.

To be concrete, and consistent with the above requirements,
consider an electron incident on the interface from the SM with energy
$\omega>X$, shown as the state $I$ of
wave vector $q^+$ in Fig.~1(a). There are four channels for
outgoing particles, labeled in Fig.~1 
by probability coefficients $A$, $B$, $C$, $D$, and by wave vectors
$q^-$, $-q^+$, $k^+$, $-k^-$, respectively.
In words, $C$ is the probability of transmission through the
interface with a wave vector on the same side of the Fermi surface (i.e.,
$q^+\rightarrow k^+$, not $-k^-$), while $D$ gives the probability
of transmission with crossing through the Fermi surface (i.e., $q^+\rightarrow
-k^-$). $B$ is the probability of intraband reflection, while
$A$ is the probability of reflection on the other side
of the Fermi surface (interband
scattering from conduction to valence band).
We write the steady state solution as
\begin{displaymath}
\Psi_{\text{SM}}(z) = \Psi_{\text{inc}}(z)+\Psi_{\text{refl}}(z),
\qquad \Psi_{\text{SC}}(z)=\Psi_{\text{trans}}(z),
\end{displaymath}
where
\begin{eqnarray}
\Psi_{\text{inc}}(z) & = &
{ 1 \choose 0 } {\rm e}^{ \text{i} q^+z },\nonumber\\
\Psi_{\text{refl}}(z) & = &
a { 0 \choose 1 } {\rm e}^{ \text{i} q^-z } +
b { 1 \choose 0 } {\rm e}^{ -\text{i} q^+z },\nonumber\\
\Psi_{\text{trans}}(z) & = &
c { u_0 \choose v_0 } {\rm e}^{ \text{i} k^+z } +
d { v_0 \choose u_0 } {\rm e}^{ -\text{i} k^-z }.
\label{eq:boundary}
\end{eqnarray}
Applying the boundary conditions, we obtain a system of four linear
equations in the four unknowns $a$, $b$, $c$, and $d$, which we solve
at a fixed value for $\omega$. We make no approximation on the
values of momenta $k^{\pm}$, $q^{\pm}$ we obtain through
Eqs.~(\ref{eq:kvect},\ref{eq:qvect}), and we introduce the
dimensionless barrier strength
\begin{displaymath}
Z=\frac{mH}{k_{\text{F}}}=H/v_{\text{F}}.
\end{displaymath}

The quantities $A$, $B$, $C$, $D$, are the ratios of the probability
current densities of the specific transmission or reflection channels
to the current of the incident particle,
e.g.~$A=\left|J_A/J_I\right|$, and so on.
The conservation of probability requires that
\begin{equation}
A + B + C + D = 1.
\label{eq:coeffsum}
\end{equation}
This result is useful in simplifying expressions for energies below
the gap, $\omega<X$, where there can be no transmitted electrons,
so that $C=D=0$. Then, Eq.~(\ref{eq:coeffsum}) reduces simply to
\protect{$A=1-B$}, so that a single quantity
is all that is needed.
Steady-state solutions other than (\ref{eq:boundary}) include an incident
electron with wave vector $-q^-$, reflected at $-q^+$ and $q^-$,
transmitted at $-k^-$ and $k^+$ ($A'$, $B'$, $C'$,
$D'$ processes, respectively), and time-reversed states for
particles incident from the SC side (with a similar set of coefficients
$A''$, $B''$, $C''$, $D''$). It can be easily shown that the primed
coefficients $A'$, $B'$, $\ldots$ are the same as $A$, $B$, $\ldots$ as well
as $A''$, $B''$, $\ldots\,$ Also, due to time-reversal symmetry,
the probability currents of the electrons coming from both the
SM and SC sides must be equal, namely
\protect{$q/m=v_{\text{g}}(u_0^2-v_0^2)$}.

\subsection{Homogeneous vs.~heterogeneous junction}\label{s:hvh}

We contrast the behavior of the homogeneous junction
with that of the heterogeneous junction by considering firstly how
below-gap reflection depends on the size of the generic gap $X$.
Kozlov and Maksimov \cite{Kozlov} studied the phase diagram
for our model of excitonic insulator. They found that, at temperature
$T=0$, the maximum value that $\Delta$ can attain is $4/\pi^2$, in units
of the effective Rydberg. The maximum is reached when, before renormalization,
the bands touch ($G=0$);
then $\Delta$ vanishes exponentially for negative values of $G$,
well inside the would-be semimetal region.
Therefore, we expect that $\Delta \ll \left|G\right|$.
Consequently, the case $X \ll \left|G\right|$ pertains to both 
SM/EI and SM/NSC junctions,
while the model can only describe the SM/NSC junction if
$X$ is larger than $\left|G\right|$.
\begin{figure*}
\centerline{\epsfig{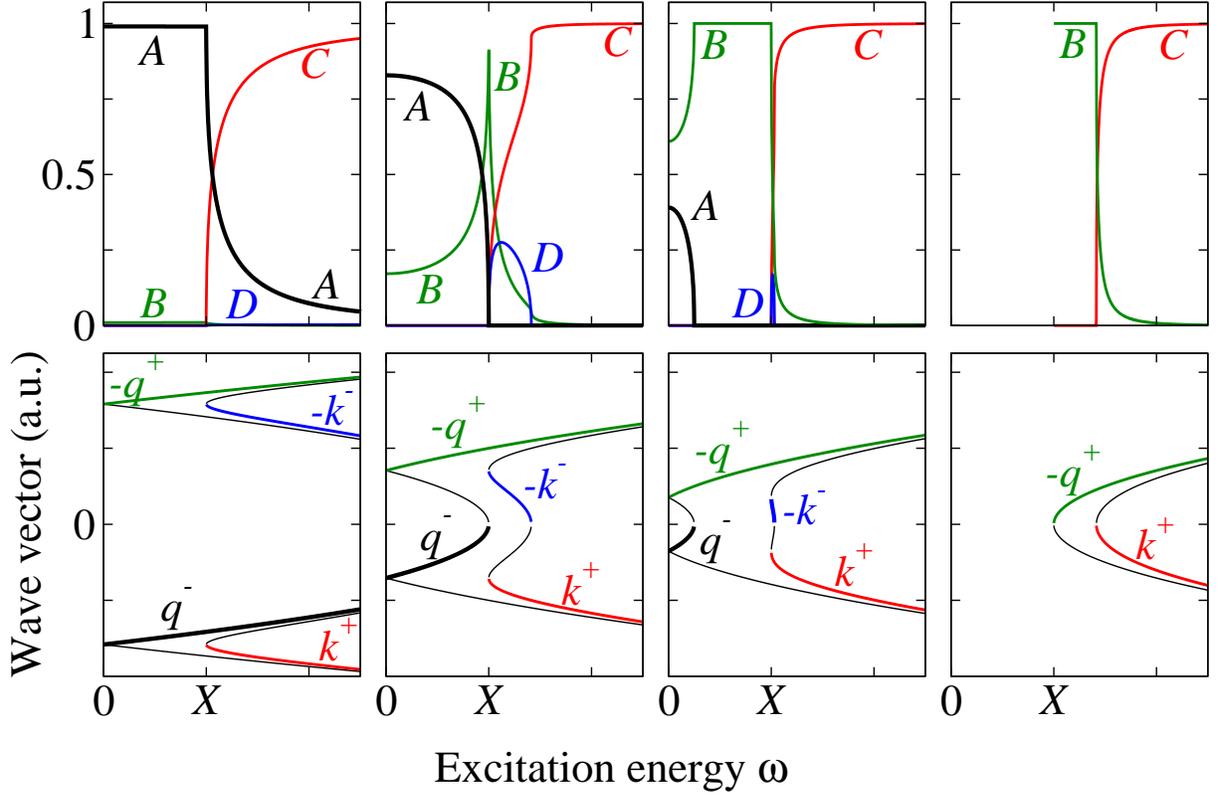}}
\caption{Band edge discontinuity at the junction.
(colour) Plot of transmission and reflection coefficients at the interface
for several values of $G/X$, as a function of the
excitation energy $\omega$ (upper row). The corresponding
quasi-particle wave vectors are also shown (lower row), labeled
as in Fig.~1.
$k$'s ($q$'s) refer to wave vectors on the SC (SM) side.
 From left to right, the four columns
correspond to $G/X=-10,-2,-0.5,2$, respectively.
We go from SM / SC ($G<0$) to SC / SC ($G>0$).
The coefficients are calculated by taking into account the exact
wave vectors of scattered particles.
$A$ gives the probability of interband (``Andreev'') reflection, $B$ gives
the probability of ordinary intraband reflection, $C$ gives the transmission
probability without branch crossing,
and $D$ transmission with branch crossing.
}
\label{fig1-si}
\end{figure*}

Figure \ref{fig1-si} shows transmission and reflection
coefficients vs.~the excitation energy $\omega$ (upper row)
for increasing values of the band gap $X$: more precisely,
we vary the ratio $G/X$ by keeping $X$ fixed and varying $G$.
We go from a wide-overlap semimetal ($G/X=-10$) up to a small-gap
semiconductor ($G/X=2$). In the latter case we allow $G$ to
be positive (SC/SC junction). In the lower row of Fig.~\ref{fig1-si}
the dispersion relations for wave vectors of transmitted
and reflected fluxes of the upper row are depicted,
in the form $q$ ($k$) vs.~$\omega$. $q$'s ($k$'s)
stand for wave vectors of carriers on the SM (SC) side:
a color code links spectrum branches (bottom row) with corresponding
probability coefficients (top row).
Data are obtained numerically,
$H_{\text{offset}}=0$, and $Z=0$.
When $X \ll \left|G\right|$ (left column, $G/X=-10$),
$A$ is almost one for $\omega<X$, then slowly decreases as
$\omega>X$. Since $B$ and $D$ are negligible,
$C\approx 1 - A$ and therefore $C$ strongly depends on $\omega$,
slowly increasing with energy as $A$ decreases.
Sub-gap interband $A$-reflection is
totally dominant and dramatically affects the transport above
the gap as well, while intraband $B$-scattering is negligible.
When $X$ increases, assuming values we
unambiguosly associate to the heterogeneous junction
($G/X=-2$, second column), the
interband  $A$-process is dramatically suppressed, both in weight and
energy range.
We note that $A$-reflection is defined only when conduction and
valence bands overlap on the SM side, namely for
\protect{$0<\omega<\left|G\right|/2$}. Here
$X=\left|G\right|/2$, and therefore one would expect, for continuity,
that the $A$-probability were suppressed limitedly to the region
where the overlap disappears, $\omega \approx \left|G\right|/2$. Instead,
$A$ is depleted over the whole allowed energy range,
$0 < \omega < \left|G\right|/2$, suggesting that
interband scattering is suppressed in the heterogeneous junction. 
Significantly, now both $B$ and $D$ 
channels are active, while
the $C$-transmission has a sharp dependence on energy only in a very
small range, above which is practically constant and close to unity.
If we sum together the weights of the two possible transmission channels,
$C$ and $D$, we see that the transmission probability is almost one
for all energies above the gap.
This is clearer at large $X$ ($G/X=-0.5$, third column):
the weight of $A$ is less than 0.5 for $\omega<\left|G\right|/2$,
while for $\left|G\right|/2<\omega<X$ the system shows
only ordinary $B$-reflection and, above the gap ($\omega>X$),
almost unit $C$-transmission. Eventually we approach the semiconductor
scenario (right column, $G/X=2$), where the $A$ channel is
absent as well as $D$ ($k^-$ is complex), and transmission
shows an almost featureless behavior, being
approximately one above the gap and zero below.
To sum up, Fig.~\ref{fig1-si} shows that, contrary to the
$X\ll \left|G\right|$ case, transmission across the 
SM/NSC junction is basically featureless and dominated by sub-gap 
intraband reflection.
\begin{figure}
\begin{picture}(210,200)
\put(-70,-50){\epsfig{file=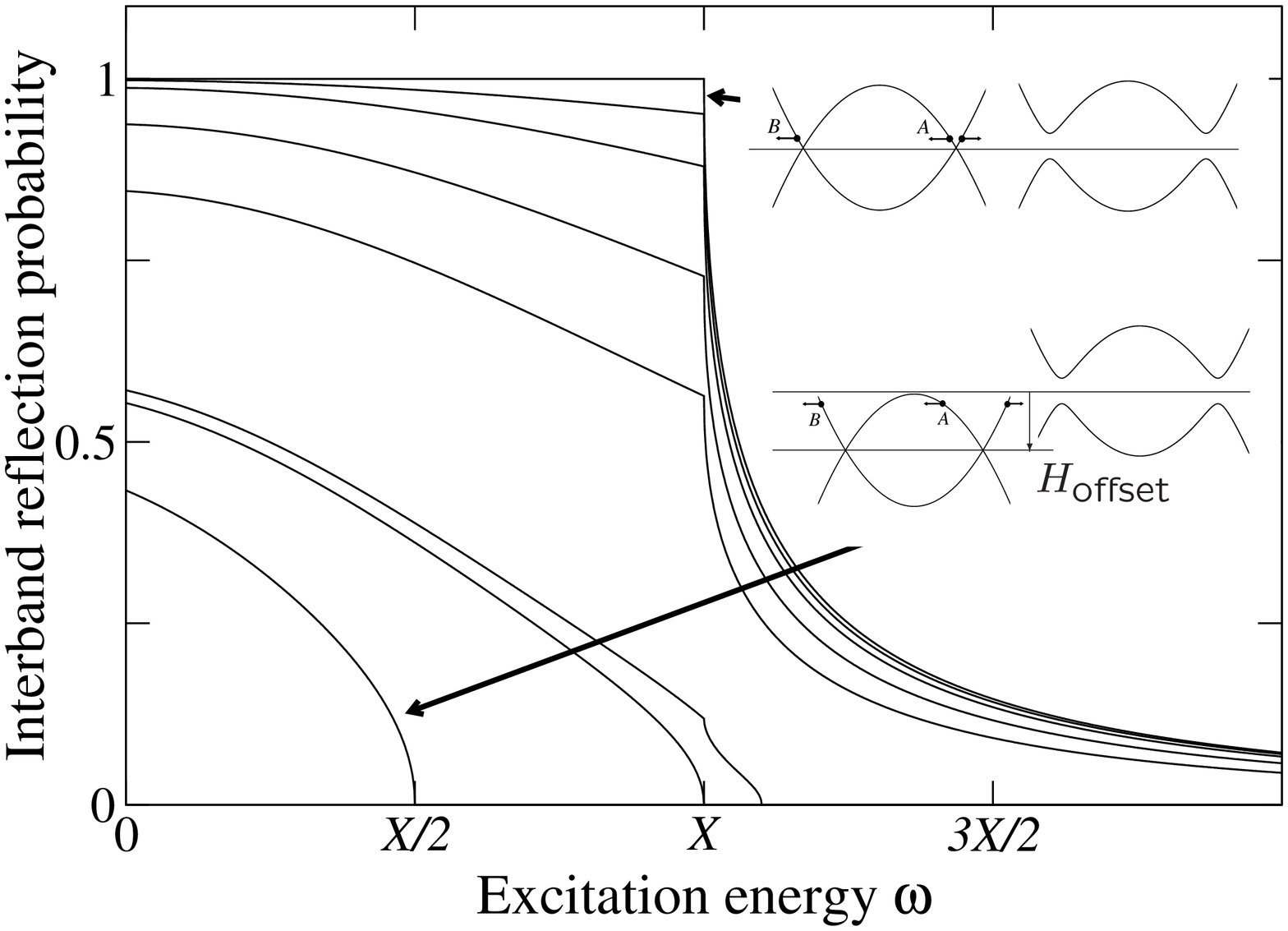,width=3.1in,,angle=90}}
\end{picture}
\caption{Short-ranged interface potential.
Plot of $A$ (interband) reflection probability
as a function of the quasiparticle energy $\omega$
for several values of the band offset $H_{\text{offset}}$.
Here $X/\left|G\right|=0.01$ and $Z=0$.
Several curves are shown from top to bottom, corresponding to
$2H_{\text{offset}}/G=0$, 0.4, 0.6, 0.8, 0.9, 0.978, 0.98, 0.99,
respectively. The two insets show the relative alignement
of SM and SC energy bands in the two cases $2H_{\text{offset}}/G=0,1$,
respectively. The $B$ (intraband) reflection probability,
for $\omega < X$, is $B=1-A$.
}
\label{fig2-si}
\end{figure}

The previous results for small values of $X$ (left column of 
Fig.~\ref{fig1-si})
are still consistent with both the homogeneous and heterogeneous cases.
Therefore, in order to differentiate the two cases,
we further consider the possibility of
band offset at the interface, i.e.~$H_{\text{offset}}\neq 0$
and $X \ll \left|G\right|$, which is solely
attributable to the heterogeneous junction.
Figure \ref{fig2-si} displays the $A$-reflection probability
vs.~$\omega$  ($X/\left|G\right|=0.01$ and $Z=0$)
for several values of the band energy offset at the interface,
$H_{\text{offset}}$. The $B$ reflection probability,
for $\omega < X$, is simply $B=1-A$.
Here we consider a negative offset, namely
SM bands are shifted downwards with respect to SC bands;
different curves from up to down correspond to
$2H_{\text{offset}}/G=0$, 0.4, 0.6, 0.8, 0.9, 0.978, 0.98, 0.99,
respectively. We see a clear suppression of the
$A$ channel as long as the energy band displacement is
increased. The two limiting cases correspond to $H_{\text{offset}}=0$,
namely there is no displacement (see top inset of Fig.~\ref{fig2-si}),
and to $2H_{\text{offset}}/G=1$, i.e.~the top of SM valence band
is exactly in line with the chemical potential $\mu=0$
(bottom inset of Fig.~\ref{fig2-si});
the latter limit is the treshold beyond which interband reflection is
forbidden. {\em A priori} one would expect strong suppression of the $A$
channel only close to the forbidden region,
\protect{$\omega \approx \left|G\right|/2
+ H_{\text{offset}}$}. Results instead show
that the $A$-weight is depleted all over the allowed
energy range. This observation is consistent with the
behavior of the heterogeneous junction for $X \approx \left|G\right|$
or larger we discussed above (Fig.~\ref{fig1-si}).
Since the band offset characterizes
the heterogeneous junction only, we conclude that
for a heterogeneous interface the $B$-channel dominates with respect to $A$.
\begin{figure}
\centerline{\epsfig{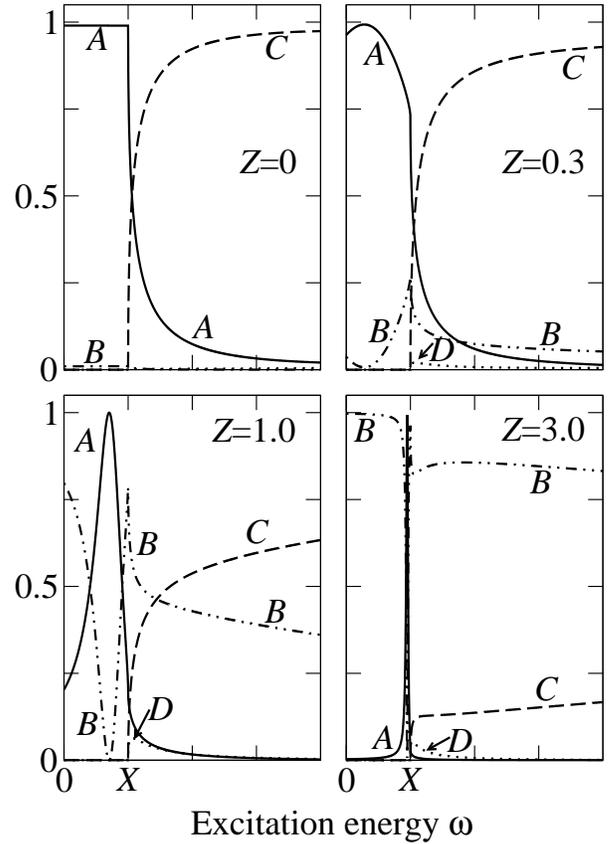}}
\caption{Effect of impurities at the interface.
Plot of transmission and reflection coefficients at the SM/SC boundary.
$A$ gives the probability of interband reflection, $B$ gives
the probability of ordinary intraband reflection, $C$ gives the transmission
probability without branch crossing, and $D$ gives the probability
of transmission with branch crossing. The parameter $Z$ measures the barrier
strength at the interface. $X/\left|G\right|=0.1$, and $H_{\text{offset}}=0$.
}
\label{fig3-si}
\end{figure}

The discussion is still incomplete, since the
appearance of a band offset is itself a hallmark
of the heterogeneous junction, but not the only one.
However, the presence of any
type of localized disorder at the interface (impurities, defects,
surface states, insulating layers, etc.)
universally characterizes the heterogeneous junction.
Therefore, we explore the case $Z\neq 0$ and $X \ll \left|G\right|$.
Figure \ref{fig3-si} shows how the dependence of transmission and
reflection coefficients on the carrier energy $\omega$
evolves as a function of the barrier strength $Z$
($X/\left|G\right|=0.1$, and $H_{\text{offset}}=0$). When the
interface is clean ($Z=0$, left top panel) almost all particles below
the gap (\protect{$\omega<X$}) are reflected with band changing,
the probability of intraband reflection being approximately zero
($A \approx 1$, $B \approx 0$).
Immediately above the gap $A$ is still very close to one,
decreasing with energy in an unexpectedly
slow manner, while $C$ increases accordingly to
$C \approx 1 - A$ ($D \approx 0$).
As disorder is added to the interface,
the probability of channel $A$ is dramatically suppressed
(Fig.~\ref{fig3-si}). Already a small amount of disorder
($Z=0.3$, top right panel of Fig.~\ref{fig3-si}) is enough to significantly
decrease the $A$-reflection probability close to the gap, while the
probabilities of $B$-reflection and $D$-transmission acquire a not
negligible value. The disorder modifies probability coefficients, 
its effect strongly depending on $\omega$ (see e.g.~plots
for $A$ and $B$ in the bottom left panel of Fig.~\ref{fig3-si}, at $Z=1$).
At $Z=3$ the interband reflection is completely suppressed, except
in a vanishingly small interval close to the gap, and the prevailing
process is the ordinary intraband $B$-reflection, whose weight is almost
uniform over the entire energy range (bottom right panel of
Fig.~\ref{fig3-si}). At high values of $Z$, the effect of
disorder is just to lower the transmission probability $C$ in an
uniform manner (therefore increasing $B \approx 1 - C$),
regardless of energy, acting as an
additional scattering source to the featureless contact resistance
of the interface. In this regime only intraband scattering is present.

We are now in the position to make a general statement on the
comparison between homogeneous and heterogeneous junctions.
We have seen in this section that in all physical cases characterizing the
heterogeneous interface, namely $X$ comparable or larger that $\left|G\right|$,
and/or presence of band offset and disorder, the $A$ interband reflection
is suppressed in favor of the $B$ intraband channel. The opposite holds for
the homogeneous junction, where the $A$ channel dominates.

\section{Junction between semimetal and excitonic
insulator\label{Andreevm}}

We now explore the physics of interband
scattering, and discard the case of the SC gap originating
from band hybridization, which is ordinary and uninteresting.
Therefore, we take $V_{\text{hyb}}=H_{\text{offset}}=0$ and
$X=\Delta$.

We generalize the model to three dimensions,
considering as an examplar case a planar interface
between a semimetal and an excitonic insulator, with
\protect{$\Delta\!\left(\bm{r},\bm{r'}\right)=\Delta\!\left(z\right)
\delta\!\left(\bm{r-r'}\right)$}, namely the pairing is a contact interaction
due to the fairly effective carrier screening in the semimetal.
Here $\Delta\!\left(z\right)$ is a smooth
increasing function of $z$, tending respectively to the asymptotic values
zero when $z\rightarrow -\infty$, well inside the bulk semimetal, and to
the constant $\Delta_0$ when $z\rightarrow +\infty$, inside the bulk
EI [Fig.~2(a)]. Let $\Delta_0$ be complex in this section.
If the characteristic scale of the spatial variation of
$\Delta\!\left(z\right)$ is much larger than the de Broglie wavelength of
carriers, the local quasi-particle dispersion relation is
\begin{equation}
\omega\!\left(\bm{k},z\right)=
\sqrt{\xi_{\bm{k}}^2+\left|\Delta\!\left(z\right)\right|^2},
\qquad \xi_{\bm{k}}=\frac{k^2}{2m}-\frac{k_{\text{F}}^2}{2m},
\end{equation}
as schematically shown in Fig.~2(a).
The quasi-particle excitations across the interface must satisfy
the equations (1) of the main text,
obtained by Eqs.~(\ref{eq:BdG}) by putting $V\!\left(\bm{r}\right)=\mu=0$.
Equations (1) of main text are formally identical
to those for bogoliubons in the intermediate state of a superconductor.
Let us now apply the method of Andreev \cite{Andreev} in
considering the reflection of quasi-particles from the boundary
separating the two phases.

The medium under consideration is completely homogeneous with an accuracy
$\left|\Delta_0/G\right|$. Therefore, since $\left|\Delta_0\right|
\ll \left|G\right|$ (cf.~App.~\ref{s:scattering}),
we seek a solution of Eqs.~(1) of main text in the form
\begin{equation}
f\!\left(\bm{r}\right)={\text{e}}^{\text{i}k_{\text{F}}\bm{n\cdot r}}
\eta\!\left(\bm{r}\right),\qquad
g\!\left(\bm{r}\right)={\text{e}}^{\text{i}k_{\text{F}}\bm{n\cdot r}}
\chi\!\left(\bm{r}\right),
\label{eq:ansatz}
\end{equation}
where $\bm{n}$ is some unit vector and $\eta\!\left(\bm{r}\right)$
and $\chi\!\left(\bm{r}\right)$ are functions that vary slowly
compared to ${\text{e}}^{\text{i}k_{\text{F}}\bm{n\cdot r}}$.
Substituting (\ref{eq:ansatz}) in (1) of main text and
neglecting higher derivatives of $\eta$ and $\chi$, we obtain
\begin{subequations}
\label{eq:BdGAndreev}
\begin{eqnarray}
\left(\text{i}v_{\text{F}}\bm{n}\cdot\nabla+\omega\right)
\eta\!\left(\bm{r}\right)-\Delta\!\left(z\right)\chi\!\left(\bm{r}\right)
&=& 0,\\
\left(\text{i}v_{\text{F}}\bm{n}\cdot\nabla-\omega\right)
\chi\!\left(\bm{r}\right)+\Delta^*\!\left(z\right)
\eta\!\left(\bm{r}\right) &=& 0,
\end{eqnarray}
\end{subequations}
where $v_{\text{F}}=k_{\text{F}}/m$.
We find the asymptotic form of the
solutions of Eqs.~(\ref{eq:BdGAndreev}) describing the reflection of
quasi-particles falling on the boundary of the semimetallic phase
when $z\rightarrow\pm\infty$.
When $z\rightarrow-\infty$ we can put $\Delta\!\left(z\right)=0$.
Then
\begin{equation}
{ \eta \choose \chi } = C_1 { 1 \choose 0 } {\text{e}}^{\text{i}
\bm{k}_1\bm{\cdot r}} + C_2 { 0 \choose 1 } {\text{e}}^{\text{i}
\bm{k}_2\bm{\cdot r}},
\label{eq:leftsolution}
\end{equation}
where ${\bm{n\cdot k}_1}=\omega/v_{\text{F}}$,
${\bm{n\cdot k}_2}=-\omega/v_{\text{F}}$; $C_1$ and $C_2$ are arbitrary
constants. The first term on the rhs of Eq.~(\ref{eq:leftsolution})
corresponds to a conduction band
electron whose velocity (or $\bm{J}$) lies along $\bm{n}$, and the second
term to a valence band electron whose velocity lies in the opposite
direction to $\bm{n}$ (in fact $\omega/v_{\text{F}}\ll k_{\text{F}}$
since \protect{$\left|\Delta_0/G\right|\ll 1$}).
If $n_z>0$, then the wavefunction (\ref{eq:leftsolution}) describes an
electron of the conduction band incident on the boundary and
reflected into the valence band on the semimetal side;
if $n_z<0$, it describes an incident
valence-band electron reflected into the conduction band.
Equations (\ref{eq:ansatz}) and (\ref{eq:leftsolution}) provide
the basis set used in Eq.~(2) of the main text.

We must put \protect{$\Delta\!\left(z\right)=\Delta_0$} in
Eqs.~(\ref{eq:BdGAndreev}) as $z\rightarrow +\infty$. The solution
describing the transmitted wave ($J_z>0$) has for
$\omega>\left|\Delta_0\right|$ the form
\begin{equation}
{ \eta \choose \chi } = \frac{C_3}{\sqrt{2}}
{
\sqrt{1 + v_{\text{F}}\bm{n\cdot k}_3/\omega}
\,{\text{e}}^{\text{i}\varphi/2}
\choose
\sqrt{1 - v_{\text{F}}\bm{n\cdot k}_3/\omega}
\,{\text{e}}^{-\text{i}\varphi/2}
}
{\text{e}}^{ \text{i}\bm{k}_3\bm{\cdot r} },
\label{eq:rightsolution}
\end{equation}
where $C_3$ is a constant, $\varphi$ is the phase of the complex
number $\Delta_0$,
\begin{subequations}
\begin{eqnarray}
{\bm{n\cdot k}_3}&=& v_{\text{F}}^{-1}
\sqrt{\omega^2-\left|\Delta_0\right|^2}\quad\text{for}\quad n_z
>  0,\\
{\bm{n\cdot k}_3}&=&- v_{\text{F}}^{-1}
\sqrt{\omega^2-\left|\Delta_0\right|^2}\quad\text{for}\quad n_z
< 0.
\end{eqnarray}
\end{subequations}
If $\omega<\left|\Delta_0\right|$, then the functions $\eta$ and
$\chi$ decay exponentially as $z\rightarrow +\infty$.

\begin{table*}
\caption{\label{tab:table2} 
Transmission through the interface.
Junction between a semimetal and an excitonic
insulator with one-dimensional equal-mass conduction and valence bands:
Transmissions and reflection coefficients. $A$ gives the probability of
``Andreev'' reflection (i.e., reflection with branch crossing), $B$ of ordinary
reflection, $C$ of transmission without branch crossing, and $D$ of
transmission with branch crossing. Here $\gamma=
Z^2\left(v_0^2-u_0^2\right)
+\left(\text{i}Z+1/2\right)2u_0^2$ and $u_0^2=1-v_0^2=1/2\,[1+(\omega^2
-\Delta^2)^{1/2}/\omega]$.}
\begin{ruledtabular}
\begin{tabular}{ccccc}
& $A$ & $B$ & $C$ & $D$ \\
\hline
No condensate ($\Delta=0$) & $0$ &
$\frac{Z^2}{1+Z^2}$ & $\frac{1}{1+Z^2}$ & $0$ \\
General form & & & & \\
$\omega < \Delta$ & $\frac{\Delta^2}{\omega^2 + 4Z^2\omega^2 +
(1+4Z^4)(\Delta^2-\omega^2)-8Z^3\omega\left(\Delta^2
-\omega^2\right)^{1/2}}$ & $1-A$ & $0$ & $0$ \\
$\omega > \Delta$ & $\frac{u_0^2v_0^2}{\left|\gamma\right|^2}$ &
$\frac{\left(u_0^2-v_0^2\right)^2Z^4+Z^2}{\left|\gamma\right|^2}$ &
$\frac{u_0^2\left(u_0^2-v_0^2\right)
\left(1+Z^2\right)}{\left|\gamma\right|^2}$ &
$\frac{v_0^2\left(u_0^2-v_0^2\right)Z^2}{\left|\gamma\right|^2}$ \\
No barrier ($Z=0$) & & & & \\
$\omega < \Delta$ & $1$ & $0$ & $0$ & $0$ \\
$\omega > \Delta$ & $\frac{v_0^2}{u_0^2}$ & $0$ &
$\frac{u_0^2-v_0^2}{u_0^2}$ & $0$ \\
Strong barrier [$Z^2(u_0^2-v_0^2)\gg 1$] & & & & \\
$\omega < \Delta$ &
$\frac{\Delta^2}{4Z^4(\Delta^2-\omega^2)}$\footnote{There is a mistake
in the corresponding result for a metal/superconductor junction
appearing in Table II of Ref.~\cite{BTK}.} & $1-A$ & $0$ & $0$ \\
$\omega > \Delta$ & $\frac{u_0^2v_0^2}{Z^4(u_0^2-v_0^2)^2}$ &
$1-\frac{1}{Z^2(u_0^2-v_0^2)}$ & $\frac{u_0^2}{Z^2(u_0^2-v_0^2)}$ &
$\frac{v_0^2}{Z^2(u_0^2-v_0^2)}$ \\
\end{tabular}
\end{ruledtabular}
\end{table*}

\subsection{Coherence factors in the
transmission coefficient}\label{anomalous}

We show that the ratio of incident electrons
$C(\omega)$ which are transmitted
through the interface [Eq.~(3) of the main text]
depends on the coherence factors
of the condensate and is strongly suppressed close to the gap.
Let us go back to the one-dimensional case and apply the boundary
conditions to wavefunctions (\ref{eq:boundary}):
we obtain an analytic solution if we let
$k^+=k^-=q^+=q^-=k_{\text{F}}$, which
is reasonable since $\Delta/\!\left|G\right|\ll 1$
(``Andreev approximation''). We find
\begin{eqnarray}
a&=& \frac{u_0v_0}{\gamma},\nonumber\\
b&=& \frac{\left(u_0^2-v_0^2\right)Z^2 -\text{i}Z}{\gamma},\nonumber\\
c&=& \frac{u_0\left(1+\text{i}Z\right)}{\gamma},\nonumber\\
d&=& -\frac{\text{i}v_0Z}{\gamma},
\end{eqnarray}
\begin{displaymath}
\gamma=Z^2\left(v_0^2-u_0^2\right)+\left(\text{i}Z+1/2\right)2u_0^2.
\end{displaymath}
In the Andreev approximation,
the probability coefficients are actually the currents, measured in
units of $v_{\text{F}}$. For example,
\protect{$A=\left|J_A\right|/v_{\text{F}}=\left|a\right|^2$}, and \protect{$D=
\left|d\right|^2/\left|v_0^2-u_0^2\right|$}.
The expression for the energy dependences of $A$, $B$, $C$, and $D$
can be conveniently written in terms of $u_0$ and $v_0$, the coherence
factors $u$ and $v$ evaluated on the branch outside the Fermi surface.
The results obtained in the Andreev approximation are given in
Table \ref{tab:table2}.
For convenience, in addition to the general
results we also list the limiting forms of the results for
zero barrier ($Z=0$) and for a strong barrier
[\protect{$Z^2(u_0^2-v_0^2)\gg 1$}], as well as for $\Delta = 0$
(the semimetal case).

Remarkably, the overall set of results of Table \ref{tab:table2}
is formally identical to
the analogous quantities obtained for a metal /
superconductor  interface
(e.g.~compare with Table II of Ref.~\cite{BTK}), the only slight
difference being the behavior for $Z\neq 0$. This is no coincidence:
the appearance of coherence factors $u$ and $v$ in probability
coefficients demonstrates that the electron-hole condensate strongly
affects the transport and in general the wavefunction of carriers,
by means of both inducing coherence on the SM side and altering
transmission features.

%
%

%


\begin{thebibliography}{30}
\bibitem{Schottky}
If a Schottky barrier is present at the interface, there is an abrupt
jump in potential and the transport behavior belongs to the
heterojunction class.
\bibitem{notecurrent}
Here, in defining the probability density $\rho$ and, consequently,
the probability current $\bm{J}$,
we consider only the intra-band free-carrier contribution.
However, we could as well have included in
the definition of $\rho$ also the inter-band contribution of the atomic
polarization: \protect{$f^*g+g^*f$}. This would add to the current $\bm{J}$ an
additional diamagnetic term whose average value
in any eigenstate of the Hamiltonian of Eq.~(\ref{eq:H})
is zero, as discussed by Nozi\`eres and Saint-James \cite{notecurrent2}.
\bibitem{SNter}
The authors define \cite{SN} the interface potential as the residual part
of the potential once that an ideal lattice discontinuity
(``flat-band condition'') has been included in the unperturbed
solution.
\end{thebibliography}
\end{document}